\documentclass[amsmath,amssymb,aps,pra,reprint,superscriptaddress]{revtex4-2}
\usepackage{graphicx}

\usepackage{color}

\begin{document}

\title{Quantum phase estimation based filtering: 
\\
performance analysis and application to low-energy spectral calculation}


\author{Rei Sakuma}
\affiliation{Materials Informatics Initiative, RD Technology \& Digital Transformation Center, JSR Corporation,
  3-103-9 Tonomachi, Kawasaki-ku, Kawasaki, 210-0821, Japan}
\affiliation{Quantum Computing Center, Keio University, 3-14-1 Hiyoshi, Kohoku-ku, Yokohama 223-8522, Japan}

\author{Kaito Wada}
\affiliation{Graduate School of Science and Technology, Keio University, 3-14-1 Hiyoshi, Kohoku, Yokohama, Kanagawa 223-8522, Japan}

\author{Shu Kanno}
\affiliation{Mitsubishi Chemical Corporation, Science \& Innovation Center, Yokohama, 227-8502, Japan}
\affiliation{Quantum Computing Center, Keio University, 3-14-1 Hiyoshi, Kohoku-ku, Yokohama 223-8522, Japan}

\author{Kimberlee Keithley}
\affiliation{Mitsubishi Chemical Corporation, Science \& Innovation Center, Yokohama, 227-8502, Japan}
\affiliation{Quantum Computing Center, Keio University, 3-14-1 Hiyoshi, Kohoku-ku, Yokohama 223-8522, Japan}

\author{Kenji Sugisaki}
\affiliation{Quantum Computing Center, Keio University, 3-14-1 Hiyoshi, Kohoku-ku, Yokohama 223-8522, Japan}
\affiliation{Graduate School of Science and Technology, Keio University, 7-1 Shinkawasaki, Saiwai-ku, Kawasaki, Kanagawa 212-0032, Japan}
\affiliation{Keio University Sustainable Quantum Artificial Intelligence Center (KSQAIC), Keio University, 2-15-45 Mita, Minato-ku, Tokyo 108-8345, Japan}
\affiliation{Centre for Quantum Engineering, Research and Education, TCG Centres for Research and Education in Science and Technology, Sector V, Salt Lake, Kolkata 700091, India}

\author{Takashi Abe}
\affiliation{Quantum Computing Center, Keio University, 3-14-1 Hiyoshi, Kohoku-ku, Yokohama 223-8522, Japan}

\author{Hajime Nakamura}
\affiliation{Quantum Computing Center, Keio University, 3-14-1 Hiyoshi, Kohoku-ku, Yokohama 223-8522, Japan}

\author{Naoki Yamamoto}
\affiliation{Quantum Computing Center, Keio University, 3-14-1 Hiyoshi, Kohoku-ku, Yokohama 223-8522, Japan}
\affiliation{Department of Applied Physics and Physico-Informatics, Keio University, 3-14-1 Hiyoshi, Kohoku-ku, Yokohama 223-8522, Japan}
\affiliation{Keio University Sustainable Quantum Artificial Intelligence Center (KSQAIC), Keio University, 2-15-45 Mita, Minato-ku, Tokyo 108-8345, Japan}

\date{\today}
\begin{abstract}
Filtering refers to the process of using algorithms to enhance desired or mask undesired states in quantum computing,
and is essential for preparing or isolating quantum many-body states.
One class of filters is based on quantum phase estimation (QPE).
Such QPE-based filters have a simple circuit implementation, but their performance remains unclear, particularly when compared to other filtering methods such as those with quantum signal processing (QSP).
 In this work, we investigate the performance
of QPE-based filters with three commonly used window functions: rectangular window, sine window, and Kaiser window.
We analyze how choice of window affects results; one noticeable observation is that
the oscillation of the filter (the Gibbs phenomenon) seen in the conventional rectangular window result can be mitigated
when the sine and Kaiser windows are applied.
Furthermore, we compare the performance of QPE-based methods to quantum eigenvalue transformation
of unitary matrices with real polynomials (QETU), and find that
the Kaiser window-based filter and QETU result in a similar number of queries to the Hamiltonian time evolution operation.
QPE-based filtering methods may therefore be an alternative to
signal-processing-based methods.
Finally, as an application of the QPE-based filter, we propose a two-step QPE
algorithm for low-energy spectral simulations, composed of a coarse grid for filtering and a fine grid
for obtaining final high-resolution spectra. As a benchmark of the proposed scheme for realistic
continuous spectra, we present the density-of-states (DOS) calculation of antiferromagnetic type-II
MnO in a one-particle approximation.
\end{abstract}

\maketitle


\section{introduction}
Calculation of electronic structure of
complex materials is one of the most promising industrial applications of quantum computing.
The last decade has seen rapid development of quantum-classical hybrid approaches that leverage currently available noisy intermediate-scale quantum computers,
such as the variational quantum eigensolver~\cite{Peruzzo2014} and
sampling-based approaches~\cite{kanno2023quantumselectedconfigurationinteractionclassical,robledomoreno2024chemistryexactsolutionsquantumcentric}.
As advances in fault-tolerant hardware and error correction techniques continue,
fully-quantum algorithms~\cite{zhang2025faulttolerantquantumalgorithmsquantum}
such as quantum phase estimation (QPE)~\cite{kitaev1995quantummeasurementsabelianstabilizer}
promise true quatum advantage in the form of exponential speed up over classical computers.

However, even with fault-tolerant hardware, preparing the precise ground state and resolving low-energy spectra are crucial,
as many phenomena of industrial and academic importance occur at the small energy scales between the electronic ground state
and low-lying excited states.
For example, semiconductor band gaps are typically on the order of $1$ eV~\cite{yu2010fundamentals}, and
the energy scale of many relevant types of chemical reactions, high-temperature superconductivity~\cite{brooks2007handbook},
and spintronics~\cite{RevModPhys.76.323,spintronics2} are in the meV range. On the
other hand, the total energy and the L2 norm of the Hamiltonian $\|H\|_{2}$ are extensive,
and even for small molecules exceed $10-100$ eV.
This difference in energy scales not only makes the ground-state preparation difficult,
as shown in recent sample-based calculations~\cite{robledomoreno2024chemistryexactsolutionsquantumcentric, D5CP02202A,yu2025quantumcentricalgorithmsamplebasedkrylov},
but also makes it hard to resolve small low-energy excitations in quantum algorithms such as QPE.

To overcome this fundamental problem, quantum filtering algorithms have been developed to obtain the desired quantum state from a given input state.
Examples of the ground-state filtering include
 the cosine~\cite{10.1063/1.5027484},
 Gaussian~\cite{PhysRevA.106.032420,Wang2022statepreparation}, and Lorentzian~\cite{Irmejs2024efficientquantum}-based filtering schemes.
There are also filtering methods based on quantum signal processing (QSP)~\cite{Low2019hamiltonian,PhysRevLett.118.010501},
such as quantum singular value transformation (QSVT)~\cite{10.1145/3313276.3316366,PRXQuantum.2.040203,Lin2020nearoptimalground} and quantum eigenvalue transformation of unitary
matrices with real polynomials (QETU)~\cite{PRXQuantum.3.040305}.
However, in all QSP-based methods one has to find
a set of single qubit phase angles to construct
the target function on a classical computer.
This procedure requires two steps. First, one needs to
find a good polynomial approximation of the target function.
Second, the phase factors realizing this polynomial function must be determined numerically.
Although classical algorithms have been developed for
the angle-finding procedure~\cite{Haah2019product,chao2020findinganglesquantumsignal,PhysRevA.103.042419,Wang2022energylandscapeof,Ying2022stablefactorization},
it is worth looking for
a simpler alternative filtering approach that avoids this procedure.

Here we choose to study filters based on QPE,
a class of algorithms that leverages the fact that eigenenergies of the system are approximately
encoded in the ancilla register in binary expansion form.
In an early work, Poulin and Wocjan~\cite{PhysRevLett.102.130503} detail
a ground-state preparation scheme based on a repeated backward operation of QPE,
and in Ref.~\cite{Jensen_2021}, P. W. K. Jense \textit{et al}. propose
a general systematic approach of amplifying quantum states within a specified energy interval.
In addition, a recent study reported~\cite{PRXQuantum.5.040339}
a simple model calculation in which their QPE-based filter with a coarse frequency grid
was more cost-effective than QETU,
suggesting QPE has the potential to outperform QSP-based methods. 
However, without
a thorough understanding of the behavior and properties of QPE-based filters,
it is difficult to assess any potential advantages.

We therefore aim to investigate the characteristics of the QPE-based filter through theoretical and
numerical analyses, particularly the effect of the initial state of the ancilla register.
This initial state, called the window function, is known to affect
the performance of QPE~\cite{PhysRevA.54.4564,PhysRevLett.98.090501,4655455,PhysRevX.8.041015,
  patel2024optimalcoherentquantumphase,greenaway2024casestudyqsvtassessment,PRXQuantum.6.020327},
prompting questions about its role in QPE-based filters.
We focus on filters that increase the probability of the states in a specified small energy region
without altering their relative amplitudes,
rather than increasing the probability of obtaining the ground state of the system as done in many ground-state preparation techniques.
This style of filtering has various applications, such as spectroscopy simulations.

We study three commonly used window functions: (1) the rectangular window, (2) the sine window, and (3) the Kaiser window.
The rectangular window is the most common and used in the conventional QPE algorithm,
although it is known to suffer from the spectral leakage problem~\cite{1455106,9762511}.
The sine window is known to mitigate the problem and has been applied in quantum metrology~\cite{PhysRevA.54.4564,PhysRevLett.98.090501,4655455,PhysRevX.8.041015}
to achieve the sought-after Heisenberg-limit scaling.
The Kaiser window~\cite{1163349} is proposed as an approximation of the prolate spheroidal window or the Slepian window~\cite{6773659,https://doi.org/10.1002/sapm196544199,Imai_2009,patel2024optimalcoherentquantumphase},
and
recently advantages of using this window over QSVT-based~\cite{PRXQuantum.2.040203,greenaway2024casestudyqsvtassessment} and
single-ancilla~\cite{PRXQuantum.6.020327} phase estimation algorithms have been reported.

First, we analyze the general properties of QPE-based filters. We show that the rectangular window-based filter exhibits an oscillation known as the Gibbs phenomenon, and the oscillation is mitigated
when the sine and Kaiser windows are used.
We also show that the effect of the window function on the filter can be incorporated into a multiplicative factor.
The error in the Kaiser window-based filter decreases exponentially with the QPE frequency grid points,
leading to the $O(\delta^{-1} \log \epsilon^{-1})$ query complexity.
Here
$\delta$ and $\epsilon$ are the width of the transitional region and the error parameter, respectively.

Next, we compare the performance of QPE-based filters and a single-ancilla QSP-based approach QETU~\cite{PRXQuantum.3.040305}.
The QETU method is known to achieve nearly optimal query complexities for the ground-state
energy estimation and the ground-state preparation~\cite{PRXQuantum.3.040305}, and 
recently a demonstration on trapped-ion quantum hardware has been reported~\cite{karacan2025filterenhancedadiabaticquantumcomputing}.
We find that the Kaiser window-based filter and QETU
with phase angles determined
via convex optimization require a comparable number of Hamiltonian evolution operations
for a given set of ($\delta, \epsilon$). This result indicates
that the Kaiser window-based filter can be used as a potential alternative to the QETU-based filter.
We discuss possible cases where the QPE-based filter can be advantageous compared to QETU.

Finally, we consider an application of the proposed filtering approach to the QPE-based calculation of spectral functions~\cite{PhysRevC.100.034610,PhysRevA.101.012330,
  PhysRevA.110.022618,PhysRevA.110.L060401,fomichev2024simulatingxrayabsorptionspectroscopy,PhysRevA.92.062318,
nishi2025demonstrationlogicalquantumphase};
specifically, a two-step QPE approach similar to the
  one proposed in Ref.~\cite{PRXQuantum.5.040339},
that uses a coarse QPE grid for filtering and a fine grid for obtaining
high-resolution spectra.
We evaluate the performance of the proposed scheme by calculating
the density-of-states (DOS) of MnO in a one-particle approximation~\cite{martin2020electronic}
as an example of realistic spectra.

\section{theory\label{sec:theory}}
\subsection{General formalism}
\begin{figure}
\includegraphics{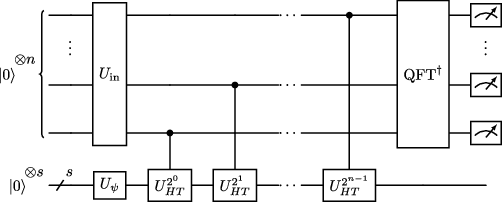}%
\caption{Quantum circuit for QPE.\label{fig:a}}
\end{figure}

Figure~\ref{fig:a} shows the general form of the quantum circuit for QPE,
which consists of $n$ and $s$ qubits for the ancilla and system registers, respectively.
Here $U_{\textrm{in}}$ constructs the input state for the ancilla register
\begin{equation}
  U_{\textrm{in}}|0\rangle^{\otimes n} = \sum_{j=0}^{N-1} a_{j}|j\rangle
  \label{eq:u_in}
\end{equation}
with $N=2^{n}$.
The real coefficients $a_{j}$ are related to the window function
in classical signal processing, and specific forms of $a_{j}$ are discussed in the next session.

The state-preparation operation $U_{\psi}$ acts on the system register as
\begin{eqnarray}
  U_{\psi}|0\rangle^{\otimes s} &=& |\psi\rangle \nonumber\\
  &=& \sum_{\mu} C_{\mu}|\phi_{\mu}\rangle,
  \label{eq:u_psi}
\end{eqnarray}
where $\phi_{\mu}$ are the eigenstates of the Hamiltonian $H$ with eigenenergies $E_{\mu}$, and $C_{\mu}$ are expansion
coefficients normalized as $\sum_{\mu} |C_{\mu}|^{2} = 1$.
After the state-preparation step, the controlled time evolution $U_{HT}=\textrm{exp}[i H T]$ is performed,
where $T$ is a time duration chosen to scale the energy range of the system in $[0, 2\pi]$.
Throughout this work $\hbar=1$.
Finally, the inverse of quantum Fourier transform (QFT) is performed.
The quantum state before the measurement in Fig.~\ref{fig:a} can be represented as
\begin{eqnarray}
  |\Psi\rangle &=& \sum_{y=0}^{N-1}
  \sum_{\mu}C_{\mu} A_{\mu}(y) |y\rangle |\phi_{\mu}\rangle,
  \label{eq:state}
\end{eqnarray}
where
\begin{equation}
  A_{\mu}(y) = \frac{1}{\sqrt{N}} \sum_{j=0}^{N-1} a_{j} \textrm{exp}[i(E_{\mu}T - \frac{2\pi}{N}y)j].
  \label{eq:amu}
\end{equation}
The probability of measuring $y=0,1,\dots,N-1$ for each state $|\phi_{\mu}\rangle$ is
\begin{equation}
  P_{\mu}(y) = |A_{\mu}(y)|^{2},
  \label{eq:pmu}
\end{equation}
and the total probability of measuring $y$ is given as the sum over $\mu$,
\begin{equation}
  P(y) = \sum_{\mu} |C_{\mu}|^{2} P_{\mu}(y).
\end{equation}
For $N \to \infty$, each $P_{\mu}(y)$ in Eq.~(\ref{eq:pmu}) has a peak at $E_{\mu}$
when seen as a function of frequency points
\begin{equation}
  \omega_{y} = \frac{2\pi}{NT}y.
  \label{eq:omega_y}
\end{equation}
However, since $P_{\mu}(y)$ is defined only on the discrete frequency points for a finite $N$,
it has, in general, nonzero tail probabilities, a phenomenon known as spectral leakage.

The QPE-based filtering amounts to discarding states with $y > y_{c}$ or $\omega_{y} > \omega_{y_{c}}$ in Eq.~(\ref{eq:state}), where $y_{c}$ is a cutoff which is taken as a positive integer. 
To see the concept clearly, we rewrite the state in Eq.~(\ref{eq:state}) as 
  \begin{eqnarray}
    |\Psi\rangle &=& \sum_{y=0}^{y_{c}} \sum_{\mu} C_{\mu} \sqrt{\mathcal{R}_{\mu}}
        \Bigl [
        \frac{A_{\mu}(y)}{\sqrt{\mathcal{R}_{\mu}}} |y\rangle
      \Bigr] |\phi_{\mu}\rangle\nonumber\\
    &&
    + \sum_{y=y_{c}+1}^{N-1}
    \sum_{\mu} C_{\mu} A_{\mu}(y)|y\rangle|\phi_{\mu}\rangle
    \nonumber\\
    &=& \sum_{\mu} C_{\mu} \sqrt{\mathcal{R}_{\mu}} |\chi_{\mu}\rangle |\phi_{\mu}\rangle
    \nonumber\\
    &&+
    \sum_{y=y_{c}+1}^{N-1}
    \sum_{\mu} C_{\mu} A_{\mu}(y)|y\rangle|\phi_{\mu}\rangle.
    \label{eq:state_to_be_filtered}
  \end{eqnarray}
Here, $\mathcal{R}_{\mu}$ is the renormalization factor, which depends on the state $|\phi_{\mu}\rangle$ through $P_{\mu}(y)$ in Eq.~(\ref{eq:pmu}) as
\begin{equation}
  \mathcal{R}_{\mu} = \sum_{y=0}^{y_{c}} P_{\mu}(y),
  \label{eq:rmu}
\end{equation}
and
  \begin{equation}
     |\chi_{\mu}\rangle = \sum_{y=0}^{y_{c}}\frac{A_{\mu}(y)}{\sqrt{\mathcal{R}_{\mu}}} |y\rangle
  \end{equation}
are normalized $\mu$-dependent states. 
Thus, the QPE-based filtering amounts to extracting the first term of Eq.~(\ref{eq:state_to_be_filtered}). 
One way to do this truncation is a coherent (i.e., fully quantum) filtering without measurement, by first performing
an ideal quantum amplitude amplification (QAA)~\cite{gilles2002,PhysRevLett.113.210501,PRXQuantum.2.040203} and then tracing out the ancilla register.
Another approach is through classical post-selection of the ancilla states after the measurement, as depicted in Fig.~\ref{fig:a}.
In this case, the result becomes probabilistic, and the success probability can be increased by performing QAA on the corresponding states.
This approach therefore differs from usual quantum filters but can be applied
to enhance the spectral estimation, as discussed in Section~\ref{sec:spectra}.

\begin{figure}
\includegraphics{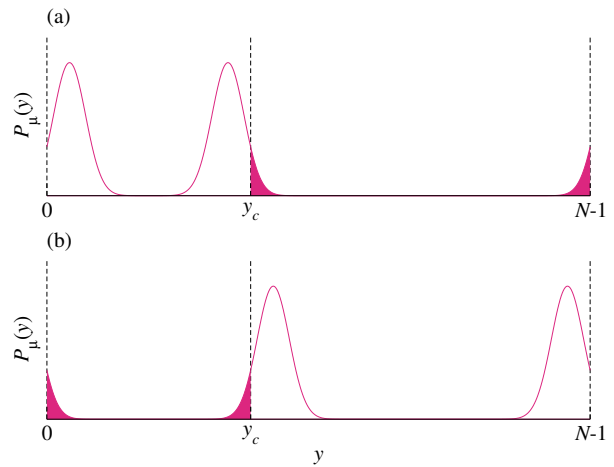}%
\caption{Errors in the QPE-based filtering. The filled regions indicate errors due to the truncation.
  (a) $P_{\mu}(y)$ for states with $E_{\mu} \leq \omega_{y_{c}}$.
  (b) $P_{\mu}(y)$ for states with $E_{\mu} > \omega_{y_{c}}$.
\label{fig:schematic}}
\end{figure}

In both quantum and classical filtering approaches, the net effect of filtering is represented as the renormalization of the weight of each state $|\phi_{\mu}\rangle$
(see the Appendix~\ref{app_sec:post_selection} for additional explanation on the classical approach):
\begin{equation}
|C_{\mu}|^{2} \to |C_{\mu}|^{2} \mathcal{R}_{\mu}.
  \label{eq:cmusqrmu}
\end{equation}
The ideal filtering corresponds to $\mathcal{R}_{\mu,\textrm{ideal}} = 1$ for the states with $0 \leq E_{\mu} \leq \omega_{y_{c}}$
and $\mathcal{R}_{\mu,\textrm{ideal}} = 0$ otherwise.
The deviation from this ideal filtering, denoted as $\Delta \mathcal{R}_{\mu}$, is given as
\begin{eqnarray}
  \Delta \mathcal{R}_{\mu} = \Bigl \{
  \begin{array}{ll}
     \sum_{y=y_{c}+1}^{N-1} P_{\mu}(y) & E_{\mu} \leq \omega_{y_{c}} \\
      \sum_{y=0}^{y_{c}} P_{\mu}(y) & E_{\mu} > \omega_{y_{c}},
    \end{array}
  \Bigr.
    \label{eq:drmu}
\end{eqnarray}
where $\sum_{y=0}^{N-1} P_{\mu}(y) = 1$ is used.
These errors originate from the tail part of the probability distribution due to spectral leakage,
as schematically shown in Fig.~\ref{fig:schematic}. Here the errors are indicated by the filled regions
in Figs.~\ref{fig:schematic}(a) and \ref{fig:schematic}(b) for states with $E_{\mu} \leq \omega_{y_{c}}$ and $E_{\mu} > \omega_{y_{c}}$,
respectively.
Note that due to periodicity the QPE-based filtering has two boundaries, at $y=y_{c}$ and at $y=0$.

Although it is not an essential point of this approach,
in this work we choose $y_{c}$ as $y_{c}=2^{n - m} - 1$ with $m(<n)$ a positive integer. With this choice, $\omega_{y_{c}}=\frac{2\pi}{NT}y_{c} \approx \frac{2\pi}{T}\frac{1}{2^{m}}$, and we
post-select or amplify the states whose first $m$ bits in the ancilla register are $0$ (ex. $|{\bf{00}}1101\rangle$ when $m=2$).
This filtering corresponds to a low-pass filter, but it is straightforward to consider filtering for general frequency intervals~\cite{Jensen_2021}.

\subsection{Window functions}

\begin{figure}
\includegraphics{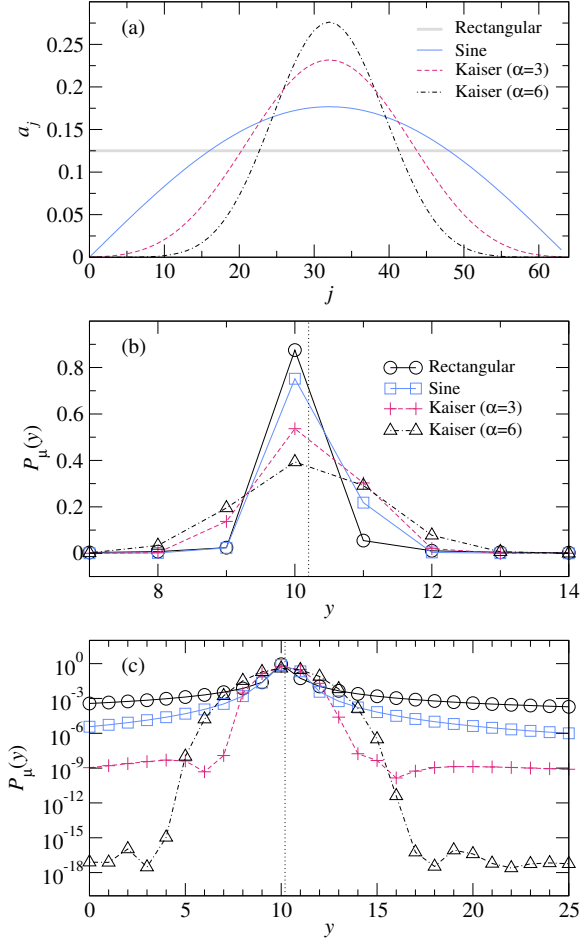}%
\caption{(a) $a_{j}$ in Eq.~(\ref{eq:u_in}) for the three windows.
  (b) $P_{\mu}(y)$ for the three windows and (c) $P_{\mu}(y)$ in the logarithmic scale.
  The vertical dotted lines in (b) and (c) indicate the position of $E_{\mu}$.
  Here $N=64$.\label{fig:a_prob}}
\end{figure}

\begin{figure}
\includegraphics{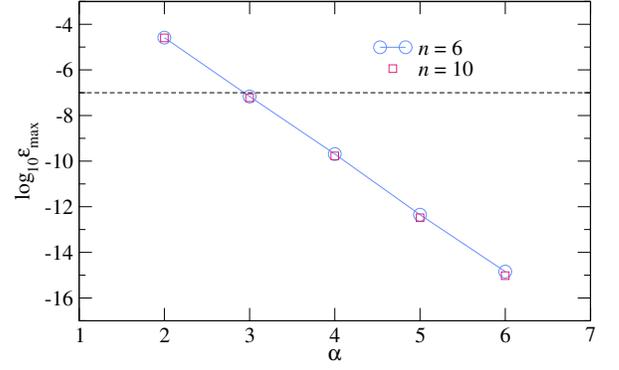}%
\caption{Error estimate in the Kaiser window-based filter defined in Eq.~(\ref{eq:eps_max}) for $n=6$ ($N=64$) and
$n=10$ ($N=1024$). The horizontal dashed line indicates $\epsilon_{\textrm{max}}=10^{-7}$ used in Sec.~\ref{sec:qetu}.
\label{fig:kaiser_eps}}
\end{figure}

The performance of the QPE-based filtering depends crucially on the form of $a_{j}$.
In this work, we consider three forms of $a_{j}$: (1) the rectangular window, (2) the sine window, and (3) the Kaiser window.
Note that many other windows have been proposed in classical signal processing~\cite{prabhu},
and a detailed exploration of their characteristics is left for future work.

The first window is
the rectangular window characterized by 
\begin{equation}
  a_{j} = \frac{1}{\sqrt{N}},
  \label{eq:a_rectangular}
\end{equation}
which is used in the conventional QPE algorithm and corresponds to setting $U_{\textrm{in}} = \textrm{H}^{\otimes n}$ with
$\textrm{H}$ the Hadamard gate. 
The root mean squared error (RMSE) of the estimated phase achieved by this window function is proportional to $1/\sqrt{N}$ known as the standard quantum limit. 

The second window is the sine window originally proposed in quantum metrology~\cite{PhysRevA.54.4564,PhysRevLett.98.090501,4655455,PhysRevX.8.041015}:
\begin{equation}
  a_{j} = \sqrt{\frac{2}{N}}\sin \frac{\pi}{N} j .
  \label{eq:a_sine}
\end{equation}
In this case, the resulting RMSE is proportional to $1/N$ \cite{4655455}, which is  known as the Heisenberg-limited scaling.
The corresponding $U_{\textrm{in}}$ can be constructed in $O(n)$ complexity~\cite{PhysRevD.106.034503}.

The third window is the Kaiser window~\cite{1163349,PRXQuantum.5.010319,greenaway2024casestudyqsvtassessment,PRXQuantum.6.020327},
which is introduced as
an approximation of the prolate spheroidal window~\cite{6773659,https://doi.org/10.1002/sapm196544199,Imai_2009,patel2024optimalcoherentquantumphase}. 
We consider the following form:
\begin{equation}
  a_{j} \propto I_{0}\left(\pi \alpha \sqrt{1 - \left(\frac{2j}{N} - 1\right)^{2}}\right),
  \label{eq:a_kaiser}
\end{equation}
where $I_{0}$ is the modified Bessel function of the order zero and $\alpha$ is a positive parameter. 
Note that our form is slightly different from other works~\cite{PRXQuantum.5.010319,greenaway2024casestudyqsvtassessment,PRXQuantum.6.020327}, in that our definition $j$ goes from $0$ to $N - 1$.
The Kaiser window and the prolate spheroidal window are more difficult to implement than the first two, but there have been proposals~\cite{mcardle2022quantumstatepreparationcoherent,patel2024optimalcoherentquantumphase}. 
In particular, the method of
  Ref.~\cite{mcardle2022quantumstatepreparationcoherent} uses
  QSVT with $O(n \sqrt[4]{\pi \alpha + 1}(\pi \alpha + \log \epsilon^{-1}))$ gate complexity within error $\epsilon$.
  In Appendix~\ref{app_sec:kaiser_qsvt},
  we review this approach and estimate the cost of constructing
  the Kaiser window.

We compare the results of all three windows in Fig.~\ref{fig:a_prob}.
In Fig.~\ref{fig:a_prob}(a) we plot $a_{j}$ for the three windows.
The Kaiser window has a Gaussian-like structure more localized than the sine window, and its spread is controlled by the parameter $\alpha$.
In Figs.~\ref{fig:a_prob}(b) and \ref{fig:a_prob}(c) we compare $P_{\mu}(y)$ defined in Eq.~(\ref{eq:pmu}) for the three windows, for a specific example of $|\phi_\mu\rangle$ with $\frac{NT}{2\pi}E_{\mu}=10.2$ and $N=64$. 
The $P_{\mu}(y)$ in the rectangular window has the highest
peak but has large leakage, and the sine window result shows a smaller tail probability~\cite{PhysRevA.110.022618}.
For these two windows, the expression of $A_{\mu}(y)$ in Eq.~(\ref{eq:amu}) can easily be obtained; for the rectangular window
\begin{eqnarray}
  A_{\mu}(y) &=& \frac{1}{N}\frac{1 - e^{i E_{\mu}T N}}{1 - e^{i \Theta_{\mu}(y)}}, 
\end{eqnarray}
and in the case of the sine window
\begin{eqnarray}
   A_{\mu}(y) &=& \frac{\sqrt{2}}{N}
   \frac{(1 + e^{i E_{\mu} T N})
     e^{i \Theta_{\mu}(y)}
      \textrm{sin}\frac{\pi}{N}}
    {(1-e^{i (\Theta_{\mu}(y) + \frac{\pi}{N})})
    (1-e^{i (\Theta_{\mu}(y) - \frac{\pi}{N})})
    },
\end{eqnarray}
where $\Theta_{\mu}(y) = E_{\mu}T - \frac{2\pi}{N}y$.
Using the inequalities
\begin{equation}
  |1 - e^{i \theta}| \leq 2
\end{equation}
and
\begin{equation}
  \frac{1}{|1 - e^{i \theta}|} \leq \frac{\pi}{2|\theta \ \textrm{mod}[-\pi, \pi)|},
\end{equation}
it can be shown that $P_{\mu}(y) = |A_{\mu}(y)|^{2}$ decays as $O(y^{-2})$ and $O(y^{-4})$ for the rectangular and
sine windows, respectively. This fact indicates that when these windows are used as a filter,
the error defined in Eq.~(\ref{eq:drmu})
also decays polynomially
in $\max \bigl\{|E_{\mu}|^{-1}, |\omega_{y_{c}} - E_{\mu}|^{-1}\bigr\}$
for $E_{\mu} \leq \omega_{y_{c}}$ or in $\textrm{max}\bigl\{|E_{\mu} - \omega_{y_{c}}|^{-1}, |\frac{2\pi}{T} - E_{\mu}|^{-1}\bigr\}$ for $E_{\mu}>\omega_{y_{c}}$.

The properties of probability distribution for the Kaiser window have been investigated in detail~\cite{PRXQuantum.5.010319,greenaway2024casestudyqsvtassessment,PRXQuantum.6.020327}.
As can be seen in Fig.~\ref{fig:a_prob}(c),
there is a clear separation between the high-probability region called mainlobe, which is confined in $\approx 2 \alpha$ QPE frequency grid points near the peak center, and the rather flat region outside the mainlobe, called sidelobes.
The sum of $P_{\mu}(y)$ in the sidelobes
gives the amount of the leakage $\epsilon$, which is related to $\alpha$ in the leading order by~\cite{PRXQuantum.5.010319}
\begin{equation}
  \alpha = O\left(\log \frac{1}{\epsilon}\right).
  \label{eq:alpha_log_epsilon_inv}
\end{equation}
To assess the error in the Kaiser window-based filter, we plot in Fig.~\ref{fig:kaiser_eps} the quantity
\begin{equation}
  \epsilon_{\textrm{max}} = \max_{E_{\mu}} \left[ 1 - \sum_{y}{}^{'} P_{\mu}(y) \right],
  \label{eq:eps_max}
\end{equation}
where the prime indicates that the sum is over the $\lceil 2 \alpha +1 \rceil$ highest-probability points near
the peak center $E_{\mu}$.
This quantity is usually larger than $\Delta \mathcal{R}_{\mu}$ in Eq.~(\ref{eq:drmu}) and
provides an estimate of the maximum error in the Kaiser window-based filter
for the states with $\textrm{min}\{E_{\mu}, \omega_{y_{c}} - E_{\mu}\} > \omega_{\lceil 2\alpha \rceil}$ or $\textrm{min}\{E_{\mu} - \omega_{y_{c}}, \frac{2\pi}{T} - E_{\mu}\} > \omega_{\lceil 2\alpha \rceil}$. 
The error decreases exponentially in $\alpha$ as expected, and the result has
a very weak dependence on the number of frequency grid points $N$.

\subsection{QPE-based filters for the three windows}

\begin{figure*}
\includegraphics{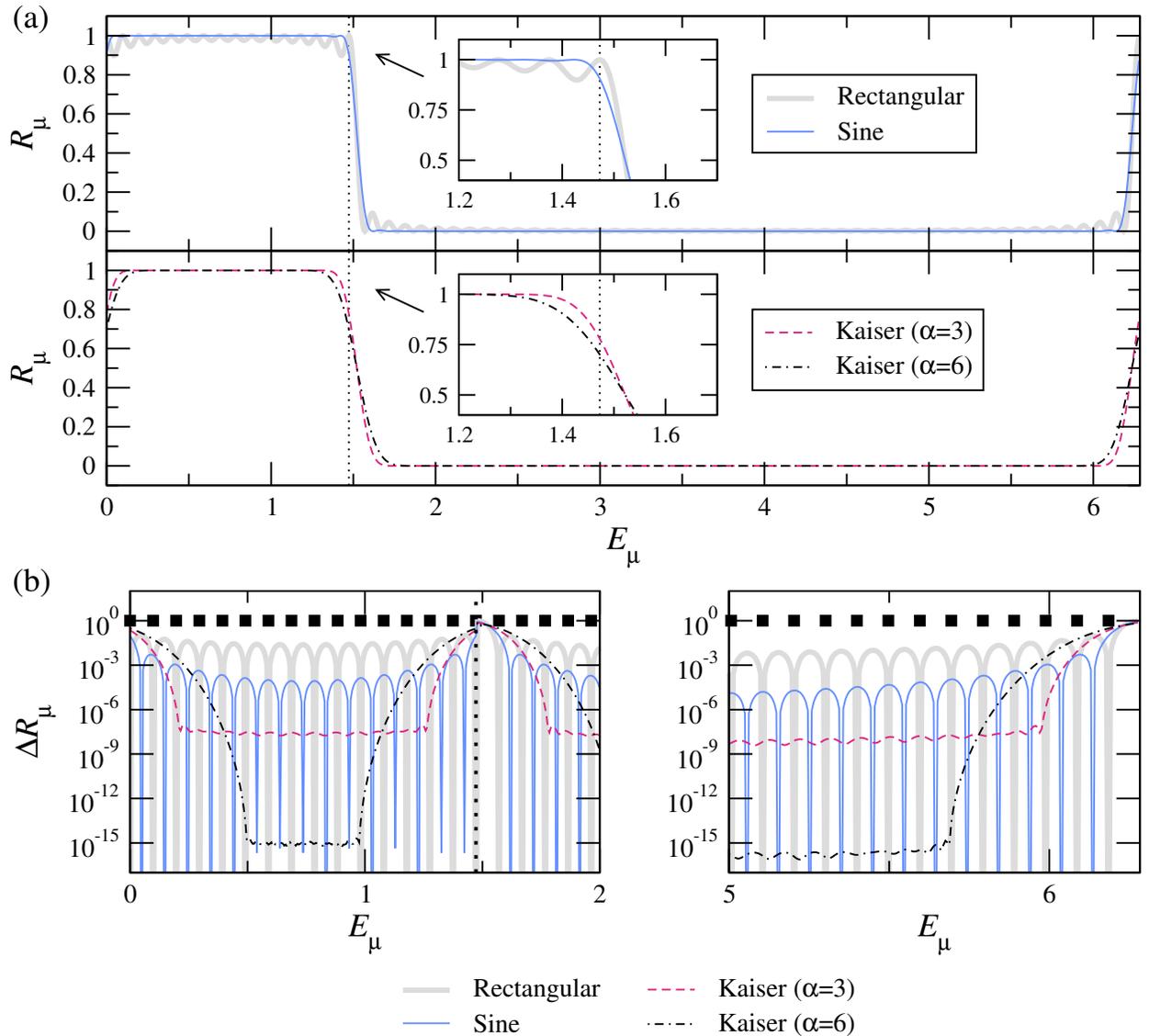}%
\caption{(a) Weight-renormalization factor (filter) $\mathcal{R}_{\mu}$ (Eq.~(\ref{eq:rmu})) for the rectangular
and the sine windows (top) and for the Kaiser window with $\alpha=3$ and $6$ (bottom).
  (b) Deviations of $\mathcal{R}_{\mu}$ from the ideal filter, $\Delta \mathcal{R}_{\mu}$ given in Eq.~(\ref{eq:drmu}), on the logarithmic scale for
  $E_{\mu} \leq 2$ (left) and $E_{\mu} \geq 5$ (right).
  The vertical dotted line in the left panel shows the position of $\omega_{y_{c}}$, and the filled squares
  indicate the positions of the QPE frequency grid points $\omega_{y}$.
  \label{fig:filter}}
\end{figure*}

Figure~\ref{fig:filter}(a) shows $\mathcal{R}_{\mu}$ (given in Eq.~(\ref{eq:rmu})) as a function of $E_{\mu}$ for the three window functions
described in the previous section. The parameters used are $T=1$, $n=6$ ($N=64$) and $y_{c}=15$, corresponding to
extracting the states in approximately
one fourth of the entire energy range. The rectangular-window-based filter exhibits an oscillating behavior
known as the Gibbs phenomenon~\cite{Jerri1998}, which originates from the discontinuity of the truncated frequency spectrum.
In the case of the other two windows this oscillation is
suppressed at this scale.
As shown in the insets of Figs.~\ref{fig:filter}(a) and \ref{fig:filter}(b),
compared to the rectangular and sine window-based filters,
the Kaiser window-based filters have a larger transitional region
where $\mathcal{R}_{\mu}$ deviates from one and zero.
This reflects the fact that the probability distribution of the Kaiser window has a broader mainlobe as shown in Figs.~\ref{fig:a_prob}(b) and \ref{fig:a_prob}(c).

In Fig.~\ref{fig:filter}(b), the errors of these filters, $\Delta \mathcal{R}_{\mu}$ defined in Eq.~(\ref{eq:drmu}), are plotted together with the QPE frequency grid points (filled squares).
The sine window result also shows oscillation, though of smaller magnitude than the rectangular window.
By analyzing $P_{\mu}(y)$~\cite{PhysRevA.110.022618}, it can be shown that these two filters are exact (i.e. $\Delta \mathcal{R}_{\mu}=0$) for $E_{\mu} = \omega_{y'}$ and $E_{\mu}=\omega_{y' + 1/2}$ for the
rectangular and sine windows, respectively, with $y'$ any integer.

The Kaiser window-based filter has the following properties:
\begin{eqnarray}
  1-\epsilon &\leq& \mathcal{R}_{\mu}, \quad \mbox{for}
  \quad E_{\mu} \in \left[\frac{\delta}{2}, E_{\textrm{targ}} + \frac{\delta}{2}\right], \label{eq:rmukaiser1}\\
  \mathcal{R}_{\mu} &\leq& \epsilon, \quad \mbox{for}  \quad E_{\mu} \in \left[E_{\textrm{targ}} + \frac{3}{2}\delta , \frac{2\pi}{T} - \frac{\delta}{2}\right].
  \label{eq:rmukaiser2}
\end{eqnarray}
Here $E_{\textrm{targ}} < \omega_{y_c}$ is the width of the target energy interval, and $\delta$ is the width of the transitional region
near $E_{\mu} =  0$, $\omega_{y_{c}}$, and $\omega_{N-1}$. The total width of the latter region is $2\delta$.
By comparing the Kaiser window results and the QPE frequency grid points,
$E_{\textrm{targ}}$ and $\delta$ can be written approximately as
\begin{eqnarray}
  \delta &\approx& \omega_{\lceil 2\alpha \rceil} = \frac{2\pi}{NT}  \lceil 2\alpha \rceil, \label{eq:delta}\\
  E_{\textrm{targ}} &\approx& \omega_{y_{c}} - \delta. \label{eq:e_targ}
\end{eqnarray}
From Eq.~(\ref{eq:delta}) and using the fact that $\alpha$ and $\epsilon$ are related as
$\alpha = O(\log \epsilon^{-1})$,
one can obtain
\begin{equation}
  N \approx \frac{2 \pi}{\delta T} 2 \alpha = O(\delta^{-1}  \log \epsilon^{-1}).
  \label{eq:n_kaiser}
\end{equation}
The corresponding number of ancilla qubits is
$O\Bigl(\log (\delta^{-1} \log \epsilon^{-1})\Bigr)$.

In practice, from a given set of ($\epsilon, \delta, E_{\textrm{targ}}$),
the parameter $\alpha$,
the number of ancilla qubits $n$,
and the cutoff parameter $y_{c}$ in the Kaiser window-based filter
can be determined as follows:
One can choose $\alpha$ from the relation between $\alpha$
and $\epsilon$ shown in Fig.~\ref{fig:kaiser_eps}.
The width of the translational region $\delta$ and $\alpha$ are related
by $\delta \approx \frac{2\pi}{2^{n}}\alpha$, from which
one can determine the number of qubits $n$.
The cutoff parameter $y_{c}$ can be selected from $2^{n}$ QPE grid points.

\subsection{The Gibbs phenomenon and the $\sigma$-factor}
As seen in the previous section, the QPE-based filter $\mathcal{R}_{\mu}$ generally
exhibits oscillation known as the Gibbs phenomenon~\cite{Jerri1998}.
To analyze this effect, let us first rewrite $\mathcal{R}_{\mu}$ as
\begin{eqnarray}
  \mathcal{R}_{\mu} = \sum_{j=-(N-1)}^{N - 1} \tilde{g}_{y_{c}}(j) \sigma_{j} e^{-i E_{\mu}T j}, 
  \label{eq:rmu2}
\end{eqnarray}
where
\begin{equation}
  \tilde{g}_{y_{c}}(j) = \frac{e^{i\frac{\pi}{N}y_{c}j}}{N} \frac{\textrm{sin}\frac{\pi (y_{c} + 1) j}{N}}{\textrm{sin}\frac{\pi j}{N}}
  \label{eq:tildegb}
\end{equation}
is the discrete Fourier transform of the discretized step function $g_{y_{c}}(y)$:
\begin{eqnarray}
  g_{y_{c}}(y) &=& \Bigl\{
  \begin{array}{ll}
    1 & \quad \mbox{for} \quad  0 \leq y \leq y_{c}\\
    0 & \quad \mbox{for} \quad y_{c} < y \leq N - 1
  \end{array}
  \Bigr. .
  \end{eqnarray}
For $N \gg j$, Eq.~(\ref{eq:tildegb}) becomes
$\tilde{g}_{y_{c}}(j) \to e^{i\frac{z_{c}}{2}j}\frac{\textrm{sin} \frac{z_{c}}{2}j}{\pi j}$ with
$z_{c} = \frac{2\pi}{N}y_{c}$,
which is the Fourier series expansion coefficient of the following $2\pi$-periodic function $G_{z_{c}}(z)$:
\begin{eqnarray}
  G_{z_{c}}(z) &=&   \Bigl\{
  \begin{array}{ll}
    1 & \quad \mbox{for} \quad 0 \leq z \leq z_{c}\\
    0 & \quad \mbox{for} \quad z_{c} < z \leq 2 \pi
  \end{array}
  \Bigr.
\end{eqnarray}
Equation~(\ref{eq:rmu2}) has the same form as the truncated Fourier series expansion of the step function with a factor $\sigma_{j}$, which is constructed from the QPE window function $a_{j}$
  \begin{equation}
    \sigma_{j} = \sum_{j'=0}^{N - 1 - |j|}  a_{|j| + j'} a_{j'}.
    \label{eq:sigma_j}
  \end{equation}
In the theory of the Gibbs phenomenon~\cite{Jerri1998}, this $\sigma$-factor is called a ``filter'' (not to be confused with filtering of states discussed in earlier sections)
and serves as a smoothing factor that improves the convergence of the truncated Fourier series.
Several filters have been proposed in the classical Fourier spectral analysis.
Some examples are the Lanczos filter and the raised cosine filter~\cite{Jerri1998,Vandeven1991}.

The explicit form of $\sigma_{j}$ for the rectangular window is
  \begin{equation}
    \sigma_{j} = \frac{N - |j|}{N}
    \label{eq:filter_rectangular}
  \end{equation}
  and for the sine window
  \begin{equation}
  \sigma_{j} = \frac{1}{N \sin\frac{\pi}{N}}
  \Bigl[
    \textrm{sin}\frac{\pi |j|}{N} \cos\frac{\pi}{N}
    +
    (N - |j|) \cos\frac{\pi |j|}{N} \sin\frac{\pi}{N}
    \Bigr].
    \label{eq:filter_sine}
    \end{equation}
  In the limit $N \to \infty$, Eqs.~(\ref{eq:filter_rectangular}) and (\ref{eq:filter_sine}) become
\begin{equation}
    \sigma(x) = 1-|x|
  \end{equation}
and
  \begin{equation}
    \sigma(x) = \frac{1}{\pi} \sin (\pi |x|) + (1-|x|) \cos (\pi x),
  \end{equation}
respectively, with $x = j/N$.

\begin{figure}
\includegraphics{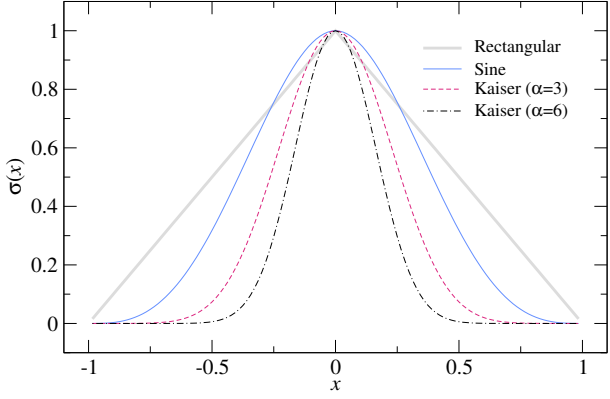}%
\caption{Comparison of the $\sigma$-factors defined in Eq.~(\ref{eq:sigma_j})
  as a function of $x = j/N$ for $N=64$.\label{fig:filters}}
\end{figure}

In Fig.~\ref{fig:filters}, we plot the above two forms as a function of $x = j/N$ as well as the Kaiser window result calculated numerically.
In the rectangular window case, $\sigma_{j}$ has a triangular structure with a singularity
at $j=0$. In the case of the sine and the Kaiser windows, $\sigma_{j}$ is more localized at $x=0$,
therefore it filters out the Fourier expansion coefficients $\tilde{g}_{y_{c}}(j)$ in Eq.~(\ref{eq:rmu2}) for large $j$.
This accelerates the convergence of the Fourier series expansion,
and yields a smoother $\mathcal{R}_{\mu}$, as observed in Fig.~\ref{fig:filter}.
More detailed analysis of the $\sigma$-factor may offer a way to improve the window function, but is beyond the scope of the present work.

\subsection{Comparison with QETU}
\label{sec:qetu}

\begin{figure*}
\includegraphics{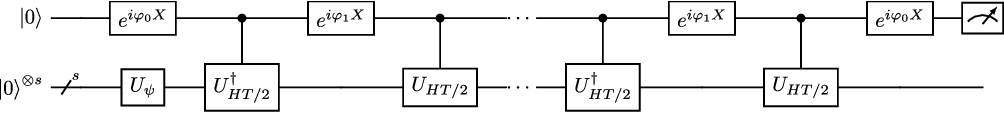}%
\caption{Quantum circuit for QETU. Here $U_{HT/2} = e^{i H T / 2}$.\label{fig:qetu_circuit}}
\end{figure*}

In this section, we compare the QPE-based and the QETU-based filters. Figure~\ref{fig:qetu_circuit} shows the quantum
circuit for QETU~\cite{PRXQuantum.3.040305} with one ancilla qubit. It consists of $d + 1$ rotation gates with
symmetric $d+1$ phase angles $(\varphi_{0},\varphi_{1},\dots,\varphi_{1}, \varphi_{0})$ and $d$ blocks of
$U^{\dagger}_{HT/2}=e^{-i H T /2}$ and $U_{HT/2}$ (note that in our case 
$HT$ is defined in the interval $[0, 2\pi]$).
The output of this circuit after measuring the ancilla qubit
is $\frac{f_{d}(w)|\psi\rangle}{||f_{d}(w)|\psi\rangle||}$,
where $f_{d}(w=\cos \frac{HT}{4})$ is an even polynomial of degree $d$.
It has been shown in Ref.~\cite{PRXQuantum.3.040305}
that the QETU-based filter can be constructed with
$d=O(\delta^{-1}\log \epsilon^{-1})$ operations of $U^{\dagger}_{HT/2}$ and $U_{HT/2}$. This is
the same order as the Kaiser window-based QPE filter, although in the case of the Kaiser window the extra cost of the window function preparation needs to be taken into account.

Our interest here is to compare the performance of the two filtering approaches in terms of the number of queries to $U_{HT}$, which has the most significant impact on the circuit depth. For this purpose, we perform a numerical simulation to
estimate the number of queries
 in both approaches for a given set of parameters $E_{\textrm{targ}}, \delta$, and $\epsilon$.
Although QETU can produce a polynomial function satisfying Eqs. (\ref{eq:rmukaiser1}) and (\ref{eq:rmukaiser2}),
here we consider a more simple and practical filtering function
$F(E)$, which corresponds to $\mathcal{R}_{\mu}$ in Eq.~(\ref{eq:rmu}), that satisfies the following conditions:
\begin{eqnarray}
  |F(E)| &\leq& 1 \quad E \in [0, \frac{2\pi}{T}] \label{eq:fecond1}\\
  1 - F(E) &\leq& \epsilon \quad E \in [0, E_{\textrm{targ}}] \label{eq:fecond2}\\
  |F(E)| &\leq& \epsilon \quad E \in [E_{\textrm{targ}} + 2 \delta, \frac{2 \pi}{T}].\label{eq:fecond3}
\end{eqnarray}
This function has the same target region $E_{\textrm{targ}}$ and the transitional region $2\delta$ as in
Eqs. (\ref{eq:rmukaiser1}) and (\ref{eq:rmukaiser2}).

Note that the filtering function $F(E)$ is applied to the squared amplitude for each state
as $|C_{\mu}|^{2} \to F(E_{\mu})|C_{\mu}|^{2}$,
rather than $C_{\mu}$ itself. We therefore
construct the function $f(w)$ such that
\begin{eqnarray}
  |f(w)| &\leq& 1 \quad w \in [0, 1] \label{eq:fscond1} \\
  f(w) &=& c \quad w \in [w_{+}, 1] \label{eq:fscond2} \\
  f(w) &=& 0 \quad w \in [0, w_{-}] \label{eq:fscond3}
\end{eqnarray}
where
\begin{eqnarray}
  c &=& \frac{1}{2}\Bigl[1 + \sqrt{1 - \epsilon} \Bigr], \label{eq:qetu_c}\\
  w_{+} &=& \cos \frac{E_{\textrm{targ}}}{4}T, \label{eq:qetu_sigma_p}\\
  w_{-} &=& \cos \frac{(E_{\textrm{targ}} + 2 \delta)}{4}T. \label{Eq:qetu_sigma_m}
\end{eqnarray}
This function is related to $F(E)$
satisfying the conditions Eqs.~(\ref{eq:fecond1}-\ref{eq:fecond3})
as $F(E) = |f(\cos\frac{ET}{4})|^{2}$.
This function is approximated by Chebyshev polynomials up to even degree $d$ as
\begin{equation}
  f(w) \approx f_{d}(w) = \sum_{l=0}^{d/2} c_{2l} T_{2l}(w),
\end{equation}
where $T_{2l}(w)$ and $c_{2l}$ are the $2l$-th Chebyshev polynomials and expansion coefficients, respectively.
We numerically determine the smallest $d$ in which $F_{d}(E)=|f_{d}(\textrm{cos}\frac{ET}{4})|^{2}$ satisfies
Eqs.~(\ref{eq:fecond1}-\ref{eq:fecond3}) for all $E \in [0, \frac{2\pi}{T}]$.

We compare two approaches to determine $c_{2l}$: (1) the polynomial approximation of the erf function
discussed by Low and Chuang~\cite{low2017hamiltoniansimulationuniformspectral}
and (2) the convex optimization approach~\cite{PRXQuantum.3.040305}.

The first approach uses
the polynomial approximation of the erf function $\textrm{erf}(k x) (k>0)$~\cite{low2017hamiltoniansimulationuniformspectral}
that is also used in the previous QPE-QETU comparison in Ref.~\cite{PRXQuantum.5.040339}:
\begin{eqnarray}
  p_{\textrm{erf},k,d+1}(x) &=& \frac{2 k e^{-k^{2}/2}}{\sqrt{\pi}}
  \Bigl(
  I_{0}(k^{2}/2) x + \nonumber\\
  &&\sum_{j=1}^{d/2} I_{j}(k^{2}/2) (-1)^{j}
  \nonumber\\
  &&
  \bigl(
  \frac{T_{2j+1}(x)}{2j+1} - \frac{T_{2j-1}(x)}{2j-1}
  \bigr)
  \Bigr).
\end{eqnarray}
The function $f_{d}(w)$ is constructed following
the approximation of the rectangular function~\cite{low2017hamiltoniansimulationuniformspectral}
\begin{eqnarray}
  f_{d}(w) &=&  c \Bigl[
    1 - \frac{1}{2}
  \Bigl(
  p_{\textrm{erf},2k,d+1}((w+w_{m})/2) \nonumber\\
    && + p_{\textrm{erf},2k,d+1}((-w+w_{m})/2)
  \Bigr)
  \Bigr] \\
  w_{m} &=& \frac{w_{+} + w_{-}}{2}\\
  k &=& \frac{\sqrt{2}}{w_{+}-w_{-}} \log^{1/2}\bigl(\frac{2}{\pi \tilde{\epsilon}^{2}}\bigr) \\
  \tilde{\epsilon} &=& 1 - \sqrt{1 - \epsilon}.
\end{eqnarray}
This $f_{d}(w)$ $\tilde{\epsilon}$-approximates $f(w)$ with
$d = O(\frac{1}{w_{+}-w_{-}} \log \tilde{\epsilon}^{-1}) = O(\frac{1}{\delta} \log \epsilon^{-1})$~\cite{low2017hamiltoniansimulationuniformspectral}.

In the second approach, the expansion coefficients $c_{2l}$ are calculated with the convex-optimization-based approach
proposed in Ref.~\cite{PRXQuantum.3.040305}
and using the \texttt{qsppack} library~\cite{PhysRevA.103.042419,Wang2022energylandscapeof,Dong2024infinitequantum}.
We use $M$ grid points $w_{m=0,1,\dots,M-1}$ in $[0, 1]$
 and solve the following optimization problem
\begin{eqnarray}
  \min_{\{c_{2l}\}}\max \Bigl[ && \max_{w_{m}\in [w_{+},1]} |f_{d}(w_{m}) - c|, \nonumber\\
  &&
    \max_{w_{m}\in [0, w_{-}]} |f_{d}(w_{m})|\Bigr] \nonumber\\
  \textrm{s.t.} && |f_{d}(w_{m})| \leq 1 \quad (m=0,1,\dots,M-1).
  \label{eq:convopt}
\end{eqnarray}

\begin{figure}
\includegraphics{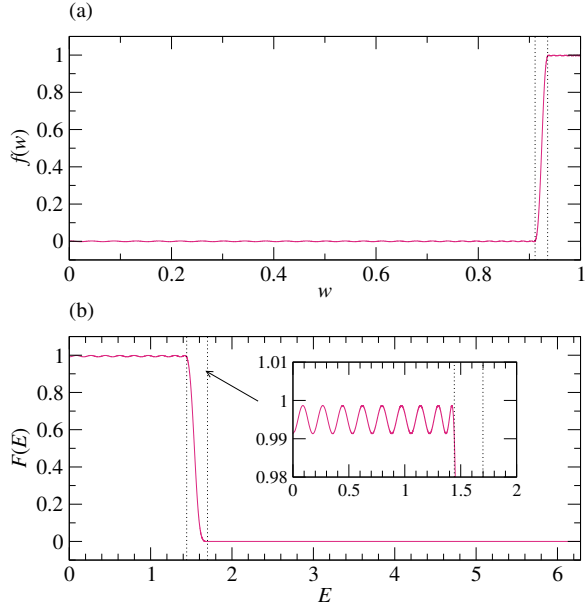}%
\caption{Polynomial approximations of (a) $f(w)$ and (b) $F(E)=|f(\cos\frac{ET}{4})|^{2}$
for $\epsilon=0.01$ calculated with the prescription of Low and Chuang~\cite{low2017hamiltoniansimulationuniformspectral}
  with
  degree 1226 and via the convex optimization~\cite{PRXQuantum.3.040305} with degree 144.
  The parameters used are $E_{\textrm{targ}}=1.4430, 2\delta=0.2556$, and $T=1$.
  The two vertical dotted lines indicate the positions of $w_{\pm}$ in (a) and
  $E_{\textrm{targ}}$ and $E_{\textrm{targ}} + 2 \delta$ in (b).
  The inset in (b) shows a magnified view.\label{fig:qetu_fpoly}}
\end{figure}

In both approaches, we check whether the function $F_{d}(E)$
satisfies Eqs.~(\ref{eq:fecond1}-\ref{eq:fecond3}) for all $E \in [0, \frac{2\pi}{T}]$ by combining
a grid search with a fine grid of $100,000$ points and an optimization routine using
the \texttt{scipy.optimization} module~\cite{2020SciPy-NMeth}.
As an example of the calculated results, $f_{d}(w)$ and $F_{d}(E)$ obtained for $\epsilon=0.01$
are shown in Figs.~\ref{fig:qetu_fpoly}(a) and \ref{fig:qetu_fpoly}(b).

\begin{figure}
\includegraphics{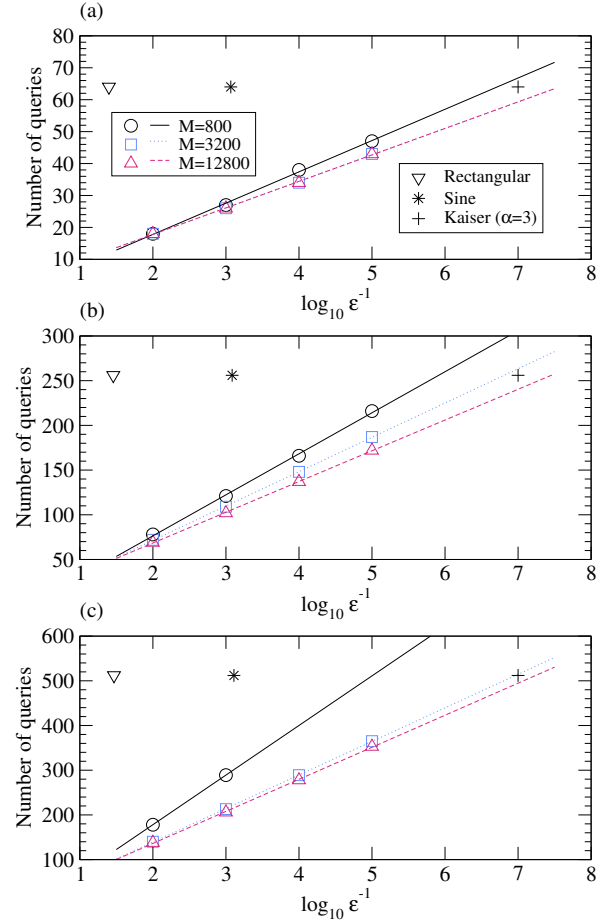}%
\caption{Minimum number of queries to $U_{HT}$ in QETU for three
sets of ($E_{\textrm{targ}}, \delta$). The corresponding values in
the QPE-based filters are also shown for comparison.
The straight lines show the results of the linear regression.
  (a) $E_{\textrm{targ}} = 1.0589$ and $ 2\delta=1.0232$.
  (b) $E_{\textrm{targ}} = 1.4430$ and $ 2\delta=0.2556$.
  (c) $E_{\textrm{targ}} = 1.5069$ and $ 2\delta=0.1279$.
  In (c) the results for $\epsilon=10^{-4}$ and $\epsilon=10^{-5}$ with $M=800$ are not calculated
as there are less sampled points than the required degree $d$.
\label{fig:qetu_query}}
\end{figure}

Figure~\ref{fig:qetu_query} shows the calculated number of queries to $U_{HT}$,
which is half of the minimum degree $d$,
for three different sets of $(E_{\textrm{targ}}, \delta)$.
Here $E_{\textrm{targ}}$ and $\delta$ are obtained numerically from the results of
the Kaiser window-based filter (Fig.~\ref{fig:filter} (b))
with $\alpha=3$ and $n=6, 8, 9$, using $y_{c}=2^{n-2} - 1$ and $\epsilon=10^{-7}$ (see also Fig.~\ref{fig:kaiser_eps}).
The number of sampled points $M$ in the second approach is a hyperparameter,
and we compare three cases, $M=800,3200$ and $12800$.
In the figures, $\epsilon$ is varied from $10^{-1}$ to $10^{-2}$ ($10^{-2}$ to $10^{-5}$) for the
first (second) approach.
The errors in the QPE-based filters with the three window functions are also shown for comparison.

It can be seen that the number of queries in the first approach is noticeably larger than
that in the second approach; for $\epsilon=10^{-2}$, queries in the former approach are $\approx 9$ times larger
than those in the latter. As a result, the QETU-based filter in the first approach requires more queries than
the QPE-based filters. This aligns with the conclusions drawn in Ref.~\cite{PRXQuantum.5.040339} based on
the rectangular window, which state that the QPE-based filtering is more efficient than QETU.

The number of required queries in the second approach, on the other hand,
is smaller than those in the QPE-based filters with the rectangular and sine windows.
It decreases as the number of sampled points $M$ increases, which
can be understood by noting that with a small $M$ the optimized coefficients
tend to overfit to the sampled data points.
Interestingly, for the three cases studied,
the QETU results in the second approach
extrapolated to $\epsilon = 10^{-7}$ are very close to the results of the Kaiser window-based filter (crosses).
Although the required queries depend on the form of the QETU filter function used, this result can partly
be understood by noting that the Kaiser window
is an approximation of the prolate spheroidal window, which is designed to maximize the probability of the mainlobe of
the probability distribution $P_{\mu}(y)$~\cite{patel2024optimalcoherentquantumphase}. This construction
is approximately equivalent to minimizing the error of the QPE-based filter $\Delta \mathcal{R}_{\mu}$ defined in Eq.~(\ref{eq:drmu}), potentially explaining the similarities between the QETU-based filter obtained from the optimization (Eq.~(\ref{eq:convopt}))
and the Kaiser window-based filter.
We note that the present calculation compares two slightly different filter forms,
but
our comparison  provides insight into the Kaiser window-based filter as an alternative to QETU and other QSP-based filtering methods.

We next discuss two situations in which the QPE-based filter could be advantageous.
The first case is when we have prior knowledge of the input spectrum.
For example, when we know the lower bound of the input spectrum $E_{\textrm{min}}$ satisfying $\delta < E_{\textrm{min}}$,
we can reduce the error $\Delta \mathcal{R}_{\mu}$
  for the states with $E_{\mu} \in [\frac{2\pi}{T} - \frac{\delta}{2}, \frac{2 \pi}{T}]$ (see Fig.~\ref{fig:filter}(b))
by shifting the Hamiltonian and the cutoff frequency as $H \to H - \frac{\delta}{2}$ and $\omega_{y_{c}} \to \omega_{y_{c}} - \frac{\delta}{2}$,
respectively.
For this shifted system, $\mathcal{R}_{\mu}$ for the Kaiser window-based filter satisfies
  \begin{eqnarray}
  1-\epsilon &\leq& \mathcal{R}_{\mu} \quad
  \quad E_{\mu}' \in \left[0, E_{\textrm{targ}} \right], \label{eq:rmukaiser1_shifted}\\
  \mathcal{R}_{\mu} &\leq& \epsilon \quad \quad E_{\mu}' \in \left[E_{\textrm{targ}} + \delta, \frac{2\pi}{T}\right],
  \label{eq:rmukaiser2_shifted}
\end{eqnarray}
where $E_{\textrm{targ}}$ is defined in Eq.~(\ref{eq:e_targ}) and $E_{\mu}'$ are shifted eigenenergies.
These two conditions are identical to the QETU ones given in Eqs.~(\ref{eq:fecond2}) and (\ref{eq:fecond3})
except that $2\delta$ in Eq.~(\ref{eq:fecond3}) is replaced by $\delta$.
Since the number of queries in QETU scales as $O(\delta^{-1} \log \epsilon^{-1})$,
this replacement approximately doubles the number of necessary queries in QETU.
Therefore, in this case
the Kaiser window-based filter is expected to be more efficient than the optimization-based QETU filter.
A similar argument can be applied
  when the upper bound of the input spectrum $E_{\textrm{max}}$ is known.
Furthermore, no shifting is required if both $E_{\textrm{min}}$ and $E_{\textrm{max}}$ are known
and satisfy $E_{\textrm{min}} > \frac{\delta}{2}$ and $E_{\textrm{max}} < \frac{2\pi}{T} - \frac{\delta}{2}$, respectively.

The second case is when a numerical optimization procedure cannot be performed as,
for example, the required accuracy is very high;
in our calculation we have difficulty obtaining reliable minimum degrees for $\epsilon < 10^{-5}$.
For very small $\delta$ and $\epsilon$ one may have to rely on simpler non-optimization-based approaches such
as Low and Chuang's procedure. In such cases, the QPE-based filter with the Kaiser window can be
a competitive alternative to QETU.

  Note that the QPE-based approach requires
  $n=O\Bigl(\log \bigl[\delta^{-1} \log \epsilon^{-1}\bigr]\Bigr)$ ancilla qubits,
  and there are additional costs from 
  the preparation of the Kaiser window as well as QFT.
  The gate complexity of QFT is $O(n^{2})$, and
  as discussed in Appendix~\ref{app_sec:kaiser_qsvt},
  the gate complexity of the QSVT-based approach in
  Ref.~\cite{mcardle2022quantumstatepreparationcoherent}
  is
  $O \Bigl(n \sqrt[4]{\pi \alpha + 1}(\pi \alpha + \log \epsilon^{-1})\Bigr) = $
  $O\Bigl(\log \Bigl[\delta^{-1} \log \epsilon^{-1}\Bigr] (\log \epsilon^{-1})^{5/4}\Bigr)$
  with three extra ancilla qubits.
  These costs are exponentially smaller than that of $N=2^{n}$ queries to
  $U_{HT}$ and $U_{HT}^{\dagger}$,
  therefore they do not become the bottleneck of
  the calculation for large systems. We also note
  that QETU requires $O(N)$ single-qubit rotations (see Fig.~\ref{fig:qetu_circuit}),
  and this cost is absent in
the QPE-based filtering.

Finally, we briefly discuss general errors in practical calculations
  for both QPE and QETU approaches, which are ignored in the simulation.
  The most important error
  comes from time propagation. Using
  Trotterization~\cite{PhysRevX.11.011020},
  the time propagation operator $U_{HT}$ is approximated as
$U_{HT}= \Bigl(\exp [i H T /r]\Bigr)^{r} \approx S_{p}(T/r)^{r}$, where
$r$ is the Trotter step, and $S_{p}(t)$ is the $p$-th
order Trotter product.
The Trotter error can be evaluated in the operator norm
\begin{equation}
  || U_{HT} - S_{p}(T/r)^{r}|| = O(C_{\textrm{Trotter}} r^{-p}),
  \label{eq:trotter}
\end{equation}
where $C_{\textrm{Trotter}}$ is a system- and $p$-dependent constant factor
constructed from nested commutators~\cite{PhysRevX.11.011020}.
A number of works have investigated tighter, more practical bounds~\cite{10.5555/2871401.2871402,PhysRevA.91.022311,Su2021nearlytight,PhysRevA.105.012403,Sahinoglu2021,Chen2024,PhysRevLett.129.270502,blunt2025montecarloapproachbound}.
With $N$ queries to $U_{HT}$ and $U_{HT}^{\dagger}$,
the required steps for precision $\epsilon_{\textrm{Trotter}}$ is
$r = O( N^{\frac{1}{p}} C_{\textrm{Trotter}}^{\frac{1}{p}}\epsilon_{\textrm{Trotter}}^{-\frac{1}{p}})$~\cite{PRXQuantum.3.040305}.
To avoid the Trotter error, 
qubitization and quantum walker based approaches have been proposed~\cite{Low2019hamiltonian,Berry2018},
and they can also be combined with the current filtering.

In the early fault-tolerant era
  the imperfectness of quantum hardware also affects the results.
Various quantum error mitigation techniques
have to been proposed to remove errors for expectation value estimation
and for sampling~\cite{RevModPhys.95.045005,liu2025quantumerrormitigationsampling},
and these techniques can be applied to obtain correct outputs.

\section{Application to calculations of low-energy spectra}
\label{sec:spectra}

\begin{figure*}
\includegraphics{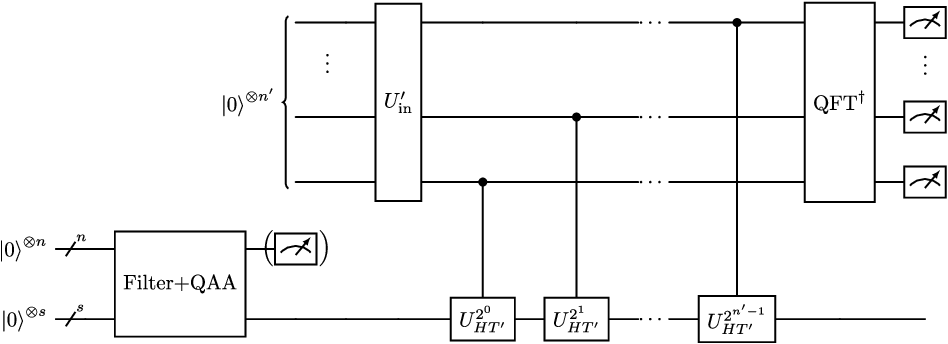}%
\caption{Quantum circuit for the QPE-based spectral simulation with filtering.\label{fig:twostep}}
\end{figure*}
\subsection{QPE-based spectral simulation}

Finally, we consider an application of the present approach to
the QPE-based calculation of the dynamical response function
\cite{PhysRevA.92.062318,PhysRevC.100.034610,PhysRevA.101.012330,
  PhysRevA.110.022618,PhysRevA.110.L060401,fomichev2024simulatingxrayabsorptionspectroscopy,
nishi2025demonstrationlogicalquantumphase}.
The objective is to calculate
\begin{equation}
  S_{B}(\omega) = \sum_{\mu} |\langle \phi_{\mu}| B| \phi_{0}\rangle|^{2} \delta(\omega - \Delta E_{\mu}),
  \label{eq:sbw}
\end{equation}
where $B$ is an operator which is not necessarily unitary, and $\Delta E_{\mu} = E_{\mu} - E_{0}$.
The quantity $S_{B}(\omega)$
is related to the Fourier transform of the dynamical response function or the correlation function
in the time domain and can be compared with various experimental spectra.

In the QPE-based approach, $S_{B}(\omega)$ is obtained using the quantum circuit in Fig.~\ref{fig:a}
by modifying the state preparation operation $U_{\psi}$ to block-encode $B|\phi_{0}\rangle$
with $a$ ancilla qubits as
\begin{eqnarray}
  \Bigl[ \langle 0 |^{\otimes a} \otimes I \Bigr] U_{\psi} |0\rangle^{\otimes a} |0\rangle^{\otimes s} &=&
  B |\phi_{0}\rangle \nonumber\\
  &=& \sum_{\mu} \langle \phi_{\mu} | B | \phi_{0} \rangle |\phi_{\mu}\rangle,
\end{eqnarray}
and by replacing $H$ in $U_{HT}$ with $H - E_{0}$ if necessary.
By collecting the measurement outputs in Fig.~\ref{fig:a},
the following approximation of $S_{B}(\omega)$ at discrete frequency points $\omega_{y}$ is obtained:
\begin{equation}
  \tilde{S}_{B}(\omega_{y}) = \sum_{\mu} |\langle \phi_{\mu}|B|\phi_{0}\rangle|^{2} P_{\mu}(y).
  \label{eq:tilde_sbw}
\end{equation}

Figure~\ref{fig:twostep} shows the quantum circuit with filtering applied prior to the QPE operation.
Such filtering is useful in cases where one is interested in some specific small energy region,
as in the case of virtual screening in materials science.
The first step of this circuit performs
filtering with $n$ ancilla qubits followed by QAA, as indicated by
the gate labeled ``Filter + QAA'' in Fig.~\ref{fig:twostep}. 
Here we apply the proposed QPE-based filtering. 
The ancilla register can either be traced out or reused after measurement.
The second step is to perform another QPE to obtain $\tilde{S}_{B}(\omega_{y})$
with $n'$ ancilla qubits and time step $T'$.
The parameters $n'$ and $T'$ as well as the window function in $U_{\textrm{in}}'$ used in
the second step
can be chosen independently to the ones used in the filtering step.

The final frequency resolution of this approach is
$\Delta\omega' = \frac{2\pi}{N' T'}$, with $N' = 2^{n'}$.
The parameters $n$ and $T$ used in the first filtering step
can be smaller than those used in the second step,
as in the coarse QPE approach proposed in Ref.~\cite{PRXQuantum.5.040339}.
In a naive application of QAA without filtering, the total number of queries to $U_{HT}$ is $O(p_{0}^{-1/2} \beta N')$,
where $\beta=T'/T$ and $p_{0}$ is the probability of measuring $\tilde{S}_{B}(\omega_{y})$ in the energy range of interest.
With filtering, the total number of queries is reduced to $O(p_{0}^{-1/2} N  + \beta N')$,
with $N = O(\delta^{-1}\log \epsilon^{-1})$ for the case of Kaiser window-based filter.
This two-step approach therefore
allows for the efficient calculation of the energy spectrum of the specified small energy region when $\beta N' \gg N$
and can be useful to get  fine structures of the spectrum at high resolution.

\subsection{Density of states of MnO}
\begin{figure}
\includegraphics{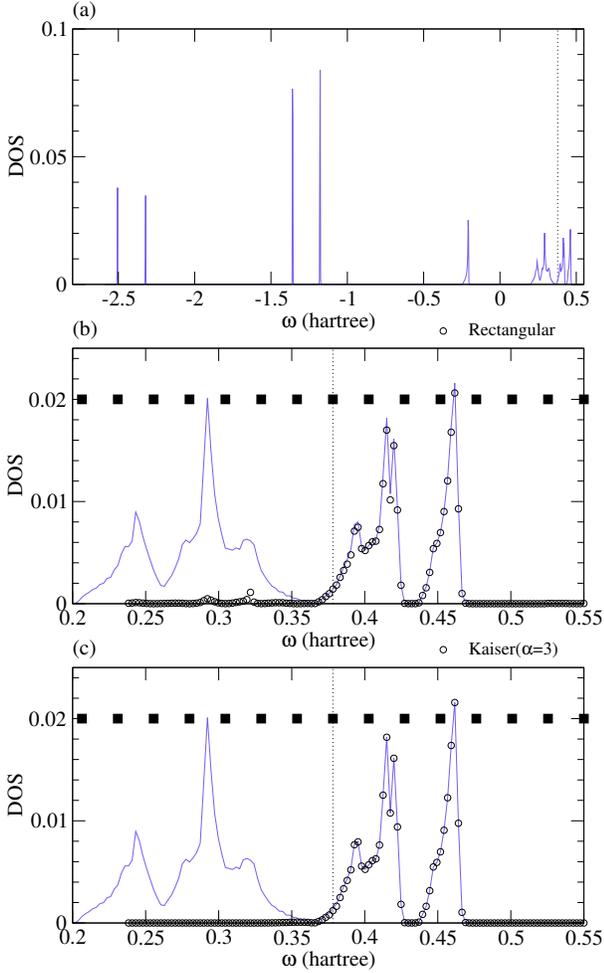}%
\caption{
  Occupied part of the DOS of MnO  (a) for the entire energy region and
  (b, c) near the cutoff after filtering using the rectangular window (b)
  and the Kaiser window with $\alpha=3$ (c).
  The solid curves show the DOS calculated with a fine QPE frequency grid, and
  the vertical dotted lines indicate the position of
  the cutoff frequency $\omega_{y_{c}}$. The states with $\varepsilon_{\mathbf{k}\nu} < \omega_{y_{c}}$ are filtered out.
  The filled squares in (b) and (c) indicate the QPE frequency points used in filtering.
  Note that the Fermi level is not set to zero in this calculation.
  \label{fig:mno_dos}}
\end{figure}

\begin{figure}
\includegraphics{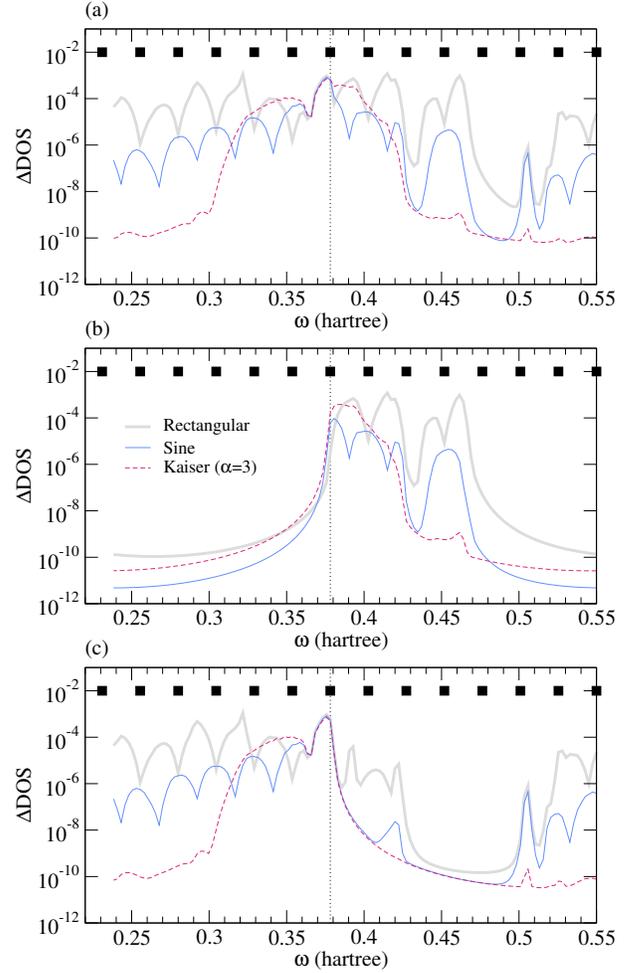}%
\caption{(a) Error of the calculated DOS after filtering.
  (b) Contribution from the states with $\varepsilon_{\mathbf{k}\nu} \geq \omega_{y_{c}}$.
  (c) Contribution from the states with $\varepsilon_{\mathbf{k}\nu} < \omega_{y_{c}}$.
  The vertical dotted lines indicate the position of $\omega_{y_{c}}$, and
  the filled squares in (b) and (c) indicate the QPE frequency points used in filtering.
  \label{fig:mno_dos_error}}
\end{figure}
To see the performance of the proposed QPE-based filtering for realistic spectra, we apply it to the calculation of the DOS~\cite{martin2020electronic} of the type-II antiferromagnetic MnO. This material
is a prototype of transition metal monoxides,
where strong electron correlation between the $3d$ orbitals of the transition metal atoms
in a narrow energy region
plays a crucial role in determining their electronic and magnetic properties.
The insulating nature of the type-II antiferromagnetic MnO is explained by
the combination of the magnetic ordering of Mn
and the coupling between Mn $3d$ and O $2p$ orbitals~\cite{PhysRevB.30.4734,PhysRevLett.52.1830},
and a number of band-structure calculations have been performed
to understand the exact electronic structure of this system and also as a benchmark of a newly proposed algorithm
~\cite{PhysRevB.51.1477,PhysRevB.76.165106,PhysRevB.78.155112,PhysRevLett.103.036404,PhysRevB.79.235114,
  PhysRevX.5.011006,PhysRevB.101.165138}.

In this work we focus on the occupied part of the DOS of this material, $D(\omega)$, within the one-particle approximation:
\begin{equation}
  D(\omega) = \sum_{\mathbf{k} \nu}^{\textrm{occupied}}
  \delta(\omega - \varepsilon_{\mathbf{k}\nu}),
  \label{eq:dos}
\end{equation}
where $\varepsilon_{\mathbf{k}\nu}$ are one-particle orbital energies,
and $\mathbf{k}$ and $\nu$ are the Bloch wavevectors
and orbital (or band) indices~\cite{martin2020electronic}, respectively.
Although this DOS itself can be calculated efficiently
  on a classical computer, we
use it as an approximation of 
the spectral function for interacting many-electron systems, which is hard to calculate classically
\cite{PhysRevA.101.012330,PhysRevA.110.L060401,PhysRevA.110.022618}.
The DOS and spectral function are related to the experimental spectra obtained in photoemission spectroscopy.

In this work we prepare $\varepsilon_{\mathbf{k}\nu}=E_{\mu=(\mathbf{k}\nu)}$
classically using density functional theory~\cite{dreizler2012density} and set
\begin{equation}
  B = \sum_{\mathbf{k} \nu}
  \hat{c}_{\mathbf{k}\nu},
\end{equation}
where $\hat{c}_{\mathbf{k}\nu}$ are
the electron annihilation operators.
The DOS is calculated via Eq.~(\ref{eq:tilde_sbw}) using a
noiseless simulator.
The input data 
is prepared
using \texttt{QuantumEspresso}~\cite{QE-2017,QE-2009}
with the pseudopotentials~\cite{martin2020electronic} 
from its pseudopotential database~\cite{QE-PPDB}.
The rocksalt structure is assumed with lattice constant $a=4.43$ \AA.
The generalized gradient approximation (GGA)~\cite{PhysRevLett.77.3865} is used
for the exchange and correlation functional, and the kinetic energy cutoff for wavefunctions used is $47$ Ry.
The one-particle energies $\varepsilon_{\mathbf{k}\nu}$ are
calculated with $32 \times 32 \times 32$
$\mathbf{k}$-points after obtaining the electron charge density self-consistently.
Although the number of states in this calculation is prohibitively large for early fault-tolerant quantum computers,
the resulting DOS yields a continuous spectrum that helps us to evaluate the efficacy of the proposed approach.

Figure~\ref{fig:mno_dos}(a) shows the DOS of this system without filtering, which is obtained with the sine window function for $U_{\textrm{in}}'$. 
The sharp peaks below $\approx -0.2$ hartree from
the Mn $3s$ and $3p$ orbitals and O $2s$ orbital play a minor role in
determining the properties of this material. The broader spectra above $\approx +0.2$
hartree
consist mainly of O $2p$ and Mn $3d$ orbitals.
Our interest here is to obtain the DOS of the Mn $3d$ orbitals at high resolution.
We therefore set a cutoff $\omega_{y_{c}}\approx 0.378$ as shown in the vertical dotted line
in Fig.~\ref{fig:mno_dos}(a), and filter out the states \emph{below} this threshold.
In the calculation, we change the sign of $\varepsilon_{\mathbf{k}\nu}$ and apply a low-pass filter without
modifying the current formalism.

The spectrum above the cutoff consists mainly of Mn $3d$ states.
As can be seen in Fig.~\ref{fig:mno_dos}(b),
due to the crystal field effect it splits into two main peaks~\cite{PhysRevLett.52.1830,PhysRevB.30.4734}:
the lower-lying $t_{2g}$ peak and the higher-lying $e_{g}$ peak.
The $e_{g}$ peak lies approximately from $0.42$ to $0.46$ hartree, and we set this energy region
as the target energy range. We add a padding of 0.5 hartree above the $e_{g}$ states.
Figures~\ref{fig:mno_dos}(b) and \ref{fig:mno_dos}(c) show the DOS (obtained via Eq.~(\ref{eq:tilde_sbw})) after QPE-based filtering in open circles,
calculated with the rectangular window
(Fig.~\ref{fig:mno_dos}(b)) and with the Kaiser window using $\alpha=3$
(Fig.~\ref{fig:mno_dos}(c)). 
Note in both cases the sine window function is used in the second step.
The parameters used are $n=n'=6$ ($N=N'=128$) and $T=2,T'=20$, and
the coarse QPE frequency grid points used in the filtering step are shown by the filled squares.

At this scale both the rectangular and the Kaiser windows efficiently
filter out the
states below the cutoff, which mainly consist of O $2p$ states,
and both results
yield higher-resolution spectra of the target ($e_{g}$) as well as $t_{2g}$ states,
in good agreement with the original DOS.
The Kaiser window result shows a better agreement with the original DOS without filtering, as expected.

Figure~\ref{fig:mno_dos_error}(a) shows the error of the calculated DOS after filtering
for the three window functions.
This error is obtained as the deviation from the unfiltered high-resolution
DOS calculated with the states above  $\omega_{y_{c}}$,
\begin{equation}
  \tilde{D}(\omega) = \sum_{\mathbf{k} \nu}^{\textrm{occupied}, \varepsilon_{\mathbf{k}\nu} \geq \omega_{y_{c}}}
  P_{\mu=(\mathbf{k}\nu)}(y).
\end{equation}
The error is largest in the rectangular window result,
and the error in the Kaiser window result decreases after $\pm \alpha$ frequency grid points used in filtering
from the cutoff frequency $\omega_{y_{c}}$, in accordance with the observation in the previous section.
Switching from the rectangular window to the Kaiser window reduces the error in the target energy region about $0.42$ to $0.46$ hartree, which corresponds to
the $e_{g}$ orbital energies (see Figs.~\ref{fig:mno_dos}(b) and \ref{fig:mno_dos}(c)), by about 
six orders of magnitude.

In Figs.~\ref{fig:mno_dos_error}(b) and \ref{fig:mno_dos_error}(c) we decompose the error into the contribution
from the states with (b) $\varepsilon_{\textbf{k}\nu} \geq \omega_{y_{c}}$ (i.e., Mn $t_{2g}$ and $e_{g}$)
and (c) $\varepsilon_{\textbf{k}\nu} < \omega_{y_{c}}$. By comparing Figs.~\ref{fig:mno_dos_error}(a) and \ref{fig:mno_dos_error}(b),
it can be seen that
the dominant error in the target region ($e_{g}$ states of Mn) originates from the states with
$\varepsilon_{\textbf{k}\nu} \geq \omega_{y_{c}}$.
Similarly, Fig.~\ref{fig:mno_dos_error}(c) shows the states with $\varepsilon_{\mathbf{k}\nu} < \omega_{y_{c}}$
are the main source of the
error in region $\omega < \omega_{y_{c}}$. In Fig.~\ref{fig:mno_dos_error}(c)
some extra low-intensity peaks are seen in the region $\omega \geq \omega_{y_{c}}$.
These peaks originate from the low-lying states (see Fig.~\ref{fig:mno_dos}(a)) whose energies
$\varepsilon_{\mathbf{k}\nu}$
are related to $\omega$ via
$\omega = \varepsilon_{\mathbf{k}\nu} + \frac{2\pi}{T'} r$ with $r$ a positive integer,
as phases in QPE are defined modulo $2\pi$.
This error is also mostly suppressed in the Kaiser window result.

In summary, this calculation demonstrates that
the proposed two-step approach provides an efficient way of calculating high-resolution spectra of the specified small energy region.
This approach is applicable to
  general discrete and continuous
  spectral simulations of quantum many-body systems,
  which are hard to calculate via classical means.
Our results show that the simplest rectangular window-based filter
already reproduces the main peak structures of the system; as the rectangular and sine windows are easy to construct,
they can prove useful when the required accuracy is not very high.
The Kaiser window-based filter provides an accurate and controllable approach, with
tunable parameters $N$ and $\alpha$ based on the specified target energy interval and the tolerance error $\epsilon$ using Eqs.~(\ref{eq:delta}) and (\ref{eq:e_targ}).

\section{conclusions}
In this work we have investigated the properties of the QPE-based filtering approach
and have shown that the choice of window functions significantly affects the quality of the results.
The results with the conventional rectangular window
exhibit an oscillation known as the Gibbs oscillation, which originates from
large spectral leakage for this window.
This oscillation is suppressed when using the sine and the Kaiser windows,
and this difference is explained by the $\sigma$-factor (smoothing factor)
appearing in the analysis of the Gibbs oscillation.
Further analysis and tuning of this factor may provide a way of improving window functions for QPE.

We have also compared the performance of the QPE-based filter and QETU in terms of necessary
queries to the time propagation operator.
We have found that, under the same conditions,
the number of queries required in the Kaiser window-based filter
is comparable to that in QETU after optimization of phase
angles. This indicates that
the form of the Kaiser window function is nearly optimal for filtering and can be used as
a potential alternative to QETU.
We have also estimated the cost of window preparation
and have shown that it is exponentially smaller than
that of the time propagation operations.
We have also discussed that
the Kaiser window-based filter can be more cost-effective than QETU
when we have prior knowledge of the energy spectrum or
filter construction in QETU is computationally hard.

As an application of the current approach,
we have considered the combination of the filtering technique with the QPE-based spectral calculation.
To assess the performance of the QPE-based filters for realistic spectra in materials science,
we have performed a numerical simulation of the DOS of type-II antiferromagnetic MnO as an approximation
of the spectral function.
Our results have shown that the QPE-based filtering efficiently extracts the states in the energy range of interest, in this case
the $e_{g}$ orbitals of Mn.
Even the simplest rectangular window sufficiently reproduces the peak structures of this system,
and the results can further be refined in a controllable manner by using the Kaiser window.
These results indicate that, combined with
efficient ground-state preparation and time evolution techniques, the proposed algorithm can be used as a simple QPE-based scheme complementary to QSP-based approaches for future materials simulations on a fault-tolerant quantum computer.

\begin{acknowledgments}
We thank Hiroshi Yano, Jumpei Kato, Kohei Oshio, and Pawan Poudel for useful discussions. 
This work was supported by MEXT Quantum Leap Flagship Program Grants No. JPMXS0118067285 and No. JPMXS0120319794. 
K.W. was supported by JSPS KAKENHI Grant Number JP24KJ1963. 
K.S. acknowledges support from Center of Innovations for Sustainable Quantum AI (JPMJPF2221) from JST, Japan, and Grants-in-Aid for Scientific Research C (21K03407) and for Transformative Research Area B (23H03819) from JSPS, Japan. 
\end{acknowledgments}

\appendix

\section{The case of classical post-selection}
\label{app_sec:post_selection}
Here we provide a detailed explanation of the classical filter
  based on post-selection.
Recall that the output state of the QPE circuit (Fig.~\ref{fig:a}) before the measurement is given by 
\begin{eqnarray*}
    |\Psi\rangle 
    &=& \sum_{y=0}^{N-1}\sum_{\mu} C_{\mu} A_\mu(y)|y\rangle |\phi_{\mu}\rangle
    \nonumber\\
    &=& \sum_{\mu} C_{\mu} \sqrt{\mathcal{R}_{\mu}} |\chi_{\mu}\rangle |\phi_{\mu}\rangle
    +\sum_{y=y_{c}+1}^{N-1}
    \sum_{\mu} C_{\mu} A_{\mu}(y)|y\rangle|\phi_{\mu}\rangle.
\end{eqnarray*}
When the measurement outcome is $y$, the state after the measurement becomes
\begin{eqnarray}
  \frac{\sum_{\mu} C_{\mu} A_{\mu}(y) |{\phi_{\mu}}\rangle}
  {\sqrt{\sum_{\mu'} |C_{\mu'}|^{2} |A_{\mu'}(y)|^{2}}},
  \label{eq:state1}
\end{eqnarray}
and the squared amplitude of state $|\phi_{\mu}\rangle$ is
\begin{equation}
  \frac{|C_{\mu}|^{2} |A_{\mu}(y)|^{2}}{\sum_{\mu'} |C_{\mu'}|^{2} |A_{\mu'}(y)|^{2}}.
  \label{eq:sqamp}
\end{equation}
The expectation value of this squared amplitude is obtained by multiplying Eq.~(\ref{eq:sqamp}) by
the conditional probability
\begin{equation}
  \frac{\sum_{\mu} |C_{\mu}|^{2} |A_{\mu}(y)|^{2}}
  {\sum_{y'=0}^{y_{c}}\sum_{\mu'} |C_{\mu'}|^{2} |A_{\mu'}(y')|^{2}}
\end{equation}
and summing over $y=0,1,\dots,y_{c}$. This cancels out the denominator of Eq.~(\ref{eq:sqamp}) and yields $|C_{\mu}|^{2}\mathcal{R}_{\mu}$ up to a constant factor. Therefore Eq.~(\ref{eq:cmusqrmu}) holds for the classical case as well
as the quantum (coherent) case.

\section{Cost of constructing the Kaiser window via QSVT}
\label{app_sec:kaiser_qsvt}

In Ref.~\cite{mcardle2022quantumstatepreparationcoherent},
a QSVT-based approach is presented
to prepare a definite-parity function $f$
using QSVT and QAA with at most three ancilla qubits.
This approach creates the operator
\begin{eqnarray}
U_{\textrm{QSVT+QAA}}
|000\rangle |0\rangle^{\otimes n} = \nonumber\\
\qquad  \mathcal{N}
\sum_{x=-\frac{N}{2}}^{\frac{N}{2} -1}
\tilde{f}(\bar{x}) |000\rangle |x\rangle ,
  \label{eq:kaiser_qsvt_qaa}
\end{eqnarray}
where $N=2^{n}$, $\bar{x}=\frac{2 x}{N}$,
$\mathcal{N} = \Bigl[\sum_{x=-\frac{N}{2}}^{\frac{N}{2}-1} |\tilde{f}(\bar{x})|^{2}\Bigr]^{-\frac{1}{2}}$,
and $\tilde{f}(\bar{x})$ is an approximation of $\frac{f(\bar{x})}{\max_{\bar{x} \in [-1, 1]} |f(\bar{x})|}$.
Note that the right hand side of
Eq.~(\ref{eq:kaiser_qsvt_qaa})
can be converted to $|000\rangle\sum_{j=0}^{N-1}a_{j}|j\rangle$ used in our window state convention
by operating a single $X$ gate.

Two building blocks are used to construct $U_{\textrm{QSVT+QAA}}$ as follows:
\begin{figure}
\includegraphics{kaiser_qsvt}
\caption{(a) Approximation of
the Kaiser window $a_{j}^{d}$
with degree $d=4$ and $12$ for $\alpha = 6$ and $n = 8$.
(b) Absolute error of
the approximated window function $\Delta a_{j} = |a_{j}^{d} - a_{j}^{\textrm{exact}}|$
for $\alpha = 6$ and $n=8$.
(c) Total number of queries to $U_{\textrm{sin}}, U_{\textrm{sin}}^{\dagger}$
in $U_{\textrm{QSVT+QAA}}$ (Eq.~(\ref{eq:kaiser_qsvt_qaa}))
and the trace distance 
$\epsilon_{td} = \sqrt{1 - (\sum_{j=0}^{N-1} a_{j}^{d} a_{j}^{\textrm{exact}})^{2}}$.
\label{fig:kaiser_qsvt}}
\end{figure}

\begin{enumerate}
\item Block-encoding of the sine function, $U_{\sin}$, with one ancilla qubit
\begin{equation}
  \Bigl( |0\rangle \langle 0| \otimes I \Bigr)
U_{\sin} |0\rangle | x\rangle
= \sin \bar{x} |0\rangle |x\rangle.
\end{equation}
This operation
requires $n + 1$ single-qubit rotations and two multitarget CNOT gates.

\item QSVT operator $U_{\tilde{f}}$ that constructs $\tilde{f}(\bar{x})$
  with another ancilla qubit
\begin{eqnarray}
\Bigl( |00\rangle \langle 00| \otimes I \Bigr)
U_{\tilde{f}}
|00\rangle |x\rangle
&=& \tilde{f}(\bar{x}) |00\rangle |x\rangle \nonumber\\
&=& h(\sin \bar{x}) |00\rangle |x\rangle .
\end{eqnarray}
Here $h(y)$
is a degree-$d$ polynomial approximation
of $\frac{f\bigl(\textrm{arcsin}(y)\bigr)}{\max_{\bar{x} \in [-1, 1]} |f(\bar{x})|}$.
This operation requires $d$ queries to $U_{\sin}$ and $U_{\sin}^{\dagger}$,
$d$ single-qubit rotations, and $2d$ CNOT gates.
\end{enumerate}

The operator $U_{\textrm{QSVT+QAA}}$ in Eq.~(\ref{eq:kaiser_qsvt_qaa})
involves $2k + 1$ operations of $U_{\tilde{f}}$ and $U_{\tilde{f}}^{\dagger}$,
$2k + 1$ single-qubit rotations, and $2k$ reflection operations.
Here $k$ is the number of applications of the QAA operator, which is
related to the L2 norm of the target state~\cite{mcardle2022quantumstatepreparationcoherent}.

In the case of the Kaiser window
\begin{equation}
  f(\bar{x}) = \frac{I_{0}(\pi \alpha \sqrt{1 - \bar{x}^{2}})}{I_{0}(\pi \alpha)},
  \label{eq:kaiser_unshifted}
\end{equation}
the overall gate complexity of this approach is
$O(n \sqrt[4]{\pi \alpha + 1}(\pi \alpha + \log \epsilon_{td}^{-1}))$~\cite{mcardle2022quantumstatepreparationcoherent},
where $\epsilon_{td}$ is the error of the resulting window state
in trace distance.
Noting that the error in the Kaiser window-based filter $\epsilon$ and
$\alpha$ are related by Eq.~(\ref{eq:alpha_log_epsilon_inv}),
by assuming $\epsilon \approx \epsilon_{td}$ we obtain
$O(n (\log \epsilon^{-1})^{\frac{5}{4}})$
as the theoretical gate complexity of the Kaiser window construction.

To evaluate the actual cost of the window construction,
we calculate optimal polynomial
coefficients of the Kaiser window for a given degree $d$
and the corresponding phase angles
using the \texttt{qsppack} library~\cite{PhysRevA.103.042419,Wang2022energylandscapeof,Dong2024infinitequantum} as done in Sec.~\ref{sec:qetu}.
In the calculation, we rescale
the target function as $f\bigl(\textrm{arcsin}(y)\bigr) \to \eta f\bigl(\textrm{arcsin}(y)\bigr)$
with $\eta = 0.98$ and use $500$ sample data points in $[0, \sin (1)]$.
This rescaling is to numerically ensure that $U_{\tilde{f}}$ is unitary
and does not affect the normalization of
the final window state~\cite{mcardle2022quantumstatepreparationcoherent}.
In Fig.~\ref{fig:kaiser_qsvt} (a) and Fig.~\ref{fig:kaiser_qsvt} (b),
we plot the approximations of the Kaiser window, $a_{j}^{d}$,
and the corresponding error $|a_{j}^{d} - a_{j}^{\textrm{exact}}|$
with $d=4$ and $12$ for $\alpha = 6$. Unlike the case
with the step function discussed in Sec.~\ref{sec:qetu},
we observe a fast convergence,
as the Kaiser function has a well-behaved expansion.

In Fig.~\ref{fig:kaiser_qsvt}(c) we plot 
the trace distance between the approximated and exact window states
$\epsilon_{td} = \sqrt{1 - (\sum_{j=0}^{N - 1} a_{j}^{d}a_{j}^{\textrm{exact}})^{2}}$
for $\alpha = 3, 6$, and $9$, as a function of
the total number of queries to $U_{\textrm{sin}}$ and $U_{\textrm{sin}}^{\dagger}$.
The total number of queries is given by the product of the degree $d$ and the
number of QAA iterations $2k+1$,
and the dominant gate cost of the calculation for large $n$ is obtained
by multiplying this number by $n + 1$~\cite{mcardle2022quantumstatepreparationcoherent}.
In the calculation $k$ is one for $\alpha = 3$
and two for $\alpha = 6$ and $9$, therefore QAA does not add a significant overhead.
In the particular case of $\alpha = 3$, 
the window state within error $\epsilon_{td} \approx 10^{-7}$, 
the same order as the error of
the Kaiser window-based filter (see Fig.~\ref{fig:kaiser_eps}),
can be obtained with $d=12$, corresponding to $36$ total queries to
$U_{\textrm{sin}}$ and $U_{\textrm{sin}}^{\dagger}$.
For $\alpha=6$ and $9$,
the number of queries seems to depend sub-linearly on $\log \epsilon_{td}^{-1}$;
our numerical fitting to $(\log \epsilon^{-1})^{\xi}$ gives $\xi \approx 0.5$.

\bibliography{refs}

\end{document}